# Multiscale study of the dynamic friction coefficient due to asperity plowing

Jianqiao Hu [a,b], Hengxu Song [c*], Stefan Sandfeld [c], Xiaoming Liu [a,b*], Yueguang Wei [d]

[a] State Key Laboratory of Nonlinear Mechanics, Institute of Mechanics, Chinese Academy of Sciences, Beijing 100190, P R China

[b] School of Engineering Science, University of Chinese Academy of Sciences, Beijing 100049, P R China

[c] Institute for Advanced Simulation, IAS-9: Materials Data Science and Informatics Forschungszentrum Juelich GmbH, Juelich 52425, Germany

[d] Department of Mechanics and Engineering Science, College of Engineering, Peking University, Beijing 100871, P R China

*Corresponding Authors: H.Song@fz-juelich.de, xiaomingliu@imech.ac.cn

**Abstract:**

A macroscopically nominal flat surface is rough at the nanoscale level and consists of nanoasperities. Therefore, the frictional properties of the macroscale-level rough surface are determined by the mechanical behaviors of nanoasperity contact pairs under shear. In this work, we first used molecular dynamics simulations to study the non-adhesive shear between single contact pairs. Subsequently, to estimate the friction coefficient of rough surfaces, we implemented the frictional behavior of a single contact pair into a Greenwood-Williamson-type statistical model. By employing the present multiscale approach, we used the size, rate, and orientation effects, which originated from nanoscale dislocation plasticity, to determine the dependence of the macroscale friction coefficient on system parameters, such as the surface roughness and separation, loading velocity, and direction. Our model predicts an unconventional dependence of the friction coefficient on the normal contact load, which has been observed in nanoscale frictional tests. Therefore, this model represents one step toward understanding some of the relevant macroscopic phenomena of surface friction at the nanoscale level.

**Keywords:**

Multiscale friction, asperity plowing, dislocation plasticity, size/velocity effect, crystal orientation, statistical model.

## 1. Introduction

The empirical mathematical summary of Amonton's first (friction force is proportional to the applied normal load) and second (friction force is independent of the apparent contact area) friction laws is the relation $F = \mu N$, where $F$ is the friction force, $N$ is the applied normal load, and $\mu$ is the coefficient of friction (COF). Even though Amonton's law works well for dry friction problems in traditional engineering, the reason for this remains unclear for quite a long time. In Bowden and Tabor's work [1], Amonton's law was explained by the fact that rough surfaces in contact with each other consist of numerous smaller contact pairs. Consequently, the actual contact area due to these microscopic and plastically deformed contact pairs is much smaller than the apparent contact area. Both experimental and numerical studies have found that the real contact area is actually proportional (or quasi-proportional) to the normal load, that is, $A = N/k$, where $k$ is a constant whose value depends on the material elasto-plastic property and surface roughness. The friction force can also be interpreted as $F = \tau A$, where $\tau$ is the shear strength of microcontacts. Therefore, one may interpret the macroscopic friction coefficient by considering the microscopic properties of the asperities as $\mu = \tau/k$.

Much attention has been dedicated to studying the linear dependence parameter $k$ between the real contact area $A$ and the normal load $N$. The Greenwood-Williamson (GW) statistical model [2] provided a convenient approach that correlates the single asperity mechanical response to rough surface contact properties. A number of studies have attempted to relax the strong assumptions and constraints of the GW model to extend its range of applicability. Examples include using the simplified elliptic model [3] for the asperity shape, considering substrate deformation to include asperity interaction [4, 5], extending the model for nearly complete contact [6], or even incorporating size-dependent plasticity [7]. The other influential theoretical approach is the Persson-type contact model [8], which implicitly includes multiple-length scales and solves the contact problem starting from the full contact condition through an analogy to the diffusion problem. Relevant works [9-14] that are based on the surface fractality also predict a linear or quasi-linear dependence between the real contact area and the normal load. The commonly accepted $k$ for elastic contact is between $E^*/\sqrt{\pi m_2/4}$ (Persson model) and $E^*/\sqrt{\pi m_2}$ (BGT model [15]), where $E^* = 1/(1-v^2)$ is the effective material modulus and $m_2$ is the second spectral moment of the rough surface. With respect to the frictional strength $\tau$ (the strength of the contact), Bowden and Tabor [1] originally claimed that it is the material shear strength. However, the estimations of the COF for metallic surfaces are not consistent with the experimentally measured values. The reason for this may be because the real contact area consists of micro-sized contact pairs, and in particular, the yield strength behaves in a size-dependent manner when an intrinsic length scale is within a range of several micrometers or less. For this reason, discrete dislocation dynamics simulations [16] have been carried out to reveal $\tau$ for different areas of contact considering the adhesive contact interface. It was found that the contact area and loading rate have a clear effect on the frictional strength, $\tau$, which is observed to have three regimes: adhesion controlled, plasticity controlled, or mixed.

It should be noted that the framework discussed above based on $k$ and $\tau$ only attributes the frictional process to the flattened asperities (Figure 1(c)). These ideally flattened asperities generally result from the contact between a rough surface and a rigid platen, and the friction corresponds to static friction. However, in a more realistic frictional model, one rough surface would slide relative to the other rough surface. At the surface asperity level, one would observe many asperity plowing pairs, as shown in Figure 1(a). The mechanical responses of these plowing asperities, at the microscopic length scales, are main reasons for macroscopic dynamic friction. Asperity plowing (Figure 1(b)) has been studied in detail, for example, by performing finite-element method (FEM) simulations [17, 18] and discrete dislocation dynamics [19, 20] and molecular dynamics (MD) simulations [21-24]. Those studies mainly focus on the asperity mechanical response under different loading conditions, such as plowing/interference depth, asperity size, and plasticity properties. Unfortunately, the connection to the surface frictional property was not considered. It can be seen in Figure 1(a) that the surface mechanical response, for example, the COF, depends on the combined average of all plowing asperities, that is, $\text{COF} = \frac{\langle T_i \rangle}{\langle N_i \rangle}$, where $T_i$, $N_i$, and $\langle \rangle$ represent the tangential force of an asperity, normal force of an asperity, and the combined average, respectively. This simple interpretation of the COF is essentially consistent with the averaging method in statistical mechanics: express the macroscopic quantities of the system at thermal equilibrium as a statistical average of microscopic functions over the canonical ensemble. Furthermore, in statistical mechanics, such a combined average can be replaced by a time average over the simulation period if, in the simulation, the fraction of time that the system spends in each state satisfies the Boltzmann distribution. For our friction problem, we assume that the rough surfaces are sufficiently large such that all pair configurations are equally likely to be found. Therefore, we can replace the combined average by the time average of a simulation where all plowing asperity configurations are covered. Such a simulation can be the plowing of a single asperity, as shown in Figure 1(b), where the deformable asperity (colored in blue) is fixed at the bottom, and the rigid asperity (colored in gray) moves rightwards under a constant velocity $V$. This simulation will contain, for example, all six configurations shown in Figure 1(a) because they have the same plowing depth. Then, the average tangential force and normal force can be written as

$$\langle T \rangle = \frac{1}{t_{cont}} \int_0^{t_{cont}} T dt, \langle N \rangle = \frac{1}{t_{cont}} \int_0^{t_{cont}} N dt, \tag{1}$$

where $t_{cont}$ represents the contact time, that is, the time duration between the instances the two asperities form contact and lose contact. Using the above idea, Mulvihill et al. [25] studied the time-averaged single asperity responses under different interference/plowing depths using FEM simulations. However, owing to the complex FEM model, which has to include large deformations and material fracture effects, the numerical convergence is difficult and the asperity interference depth in their FEM modeling is quite small. Furthermore, there is also limited plastic deformation in the plowed asperity. In addition, the application of conventional continuum plasticity for surface asperity studies is always debatable as it is well known that for crystalline materials, plasticity becomes size dependent at such small length scales. Therefore, it is necessary to utilize simulation techniques that can capture material elastoplastic properties at a small length scale, such as MD simulations.

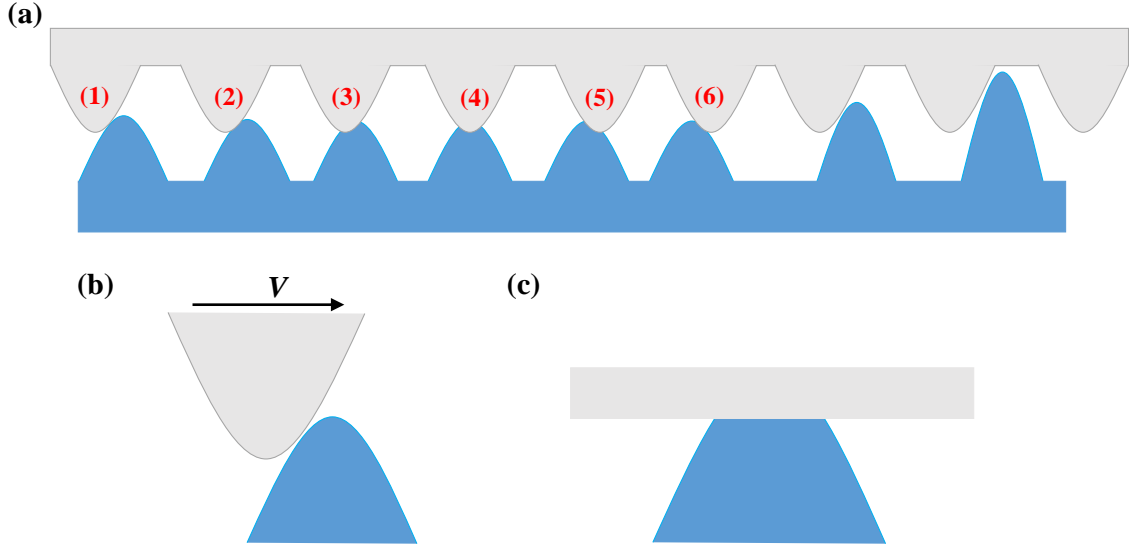

Figure 1: (a) Schematic of rough surfaces during friction: the gray surface is rigid while the blue surface is deformable. For plowing asperity pairs that have the same plowing depth (distance between the asperity tips), they can form different contact configurations indicated by the numbers. Plowing pairs can also have different plowing depths represented by the last two plowing pairs. (b) Schematic of single asperity plowing problem. The deformable asperity (blue) is fixed at the bottom while the rigid asperity (gray) moves rightwards under a constant velocity $V$. (c) Simplified asperity flattening configuration, which is a special configuration of asperity plowing, i.e., equivalent to configuration (4) in (a).

A significant body of literature exists concerning friction-related studies using MD simulations. For example, for a nanoscratching problem, Zhang and Tanaka [26] revealed that there are generally four distinct regimes of deformation during a diamond-copper sliding system, that is, no-wear, adhering, plowing, and cutting regimes. The dominant regimes are governed by some sliding parameters such as invasion depth, sliding velocity, and surface lubrication conditions. Gunkelmann et al. [27] carried out MD simulations of nanoscratching on iron, and the dislocation microstructure was characterized using continuum field quantities from continuous dislocation dynamics theory [28]. The mechanical response of the material is then linked to the evolution of the dislocation field quantities. For grinding processes, the effect of the sliding velocity on the subsurface damage was investigated by analyzing the dislocation and phase transformation processes [29]. It was proposed that the subsurface damage thickness only slightly increased when the sliding velocity exceeded 180 m/s. With respect to the friction law at the nanometer length scale, Müser et al. [30] established a microscopic theory to explain the molecular origins of friction. MD simulations were then used to verify their theoretical predictions. Mo et al. [31] studied the dependence of the friction force on the applied load and contact area at a nanoscale level, and they demonstrated that the friction force is linearly related to the number of interacting atoms.

To the best of our knowledge, the connection between single asperity plowing responses at the atomistic scale and the rough surface frictional property is missing. Herein, we perform a multiscale study. For the small length scale (i.e., asperity scale), we study in detail the single asperity response during the whole plowing process by performing MD simulations. The effects of interference depth, asperity size, sliding velocity, and crystal orientation are considered. For the large length scale (surface length scale), the single asperity response is utilized to predict the COF during surface sliding through a GW-type statistical modeling. The remainder of this paper is organized as follows. Section 2 presents the methodology and model description. Next, the MD simulation results of single asperity plowing are discussed in Section 3. Section 4 focuses on the prediction of the COF based on statistical modeling. A discussion and concluding remarks are given in Section 5. The effects of crystalline orientation and numerical fitting of the asperity responses are discussed in Appendices A and B, respectively.

## 2. Methodology and model description

The large-scale atomic/molecular massively parallel simulator [32] is employed for MD simulations. The geometry of the simulation model is illustrated in Figure 2. A hemispherical asperity with the face-centered cubic copper is located at the center of a copper cubic substrate. The radius $R$ of the hemisphere varies from $5a$ to $50a$, with $a$ representing the lattice constant of 3.615 Å. The lengths of the substrate are $4R$ in the $x$ and $y$ directions (i.e., $l_x$ and $l_y$), and the thickness $l_z$ is $0.8R$. The size of the substrate is large enough such that no dislocations can reach the lateral and bottom boundaries during the entire plowing process. Plowing is conducted in a displacement controlled manner by moving the upper rigid asperity of the same size along the $y$ direction under a constant velocity $V$, while the bottom of the substrate is fixed. The interference/plowing depth is chosen as $h = \alpha \cdot R$.

The interaction between asperities is simulated by a repulsive potential with a force of magnitude

$$F = \begin{cases} -K(r_i - R)^2 & \text{if } r_i \leq R \\ 0 & \text{otherwise} \end{cases} \quad (2)$$

Here, $K$ is a constant that is related to the effective stiffness of the rigid asperity indenter, and is set to 10 eV/Å$^3$, $R$ is the radius of the indenter, and $r_i$ is the distance from the $i$th atom to the center of the rigid asperity (Figure 2 only shows its lower half). Adhesion at the contact interface plays an important role during nanoscale contact; however, in this study, we focused on dislocation plasticity; therefore, we excluded the adhesion. The repulsive potential eliminates the effect of interface adhesion and the tangential interaction at the contact interface, which is equivalent to having a frictionless interface.

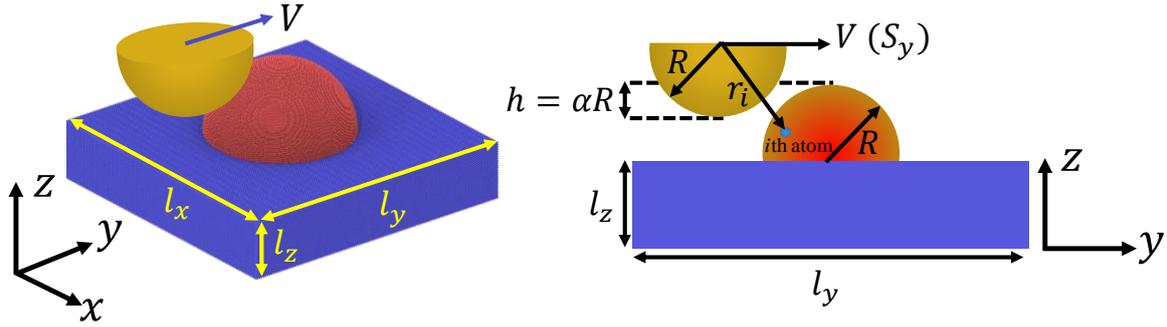

Figure 2: Schematic of spherical asperity plowing in the MD simulation.

In the simulations, the $x$, $y$, and $z$-axes were initially chosen as the [001], [010], and [001] lattice directions, respectively. The mechanical properties of the asperity also depend on the crystal orientations; a detailed discussion can be found in Appendix A. Periodic boundary conditions are imposed along the $x$ and $y$ directions, while the $z$ direction is non-periodic. The interactions among copper atoms in the asperities are described by the embedded atom method (EAM) potential [33], which has been widely used in the simulation of copper [34-37]. Prior to the movement of the rigid asperity, the initial structures are fully relaxed at 0.01 K, with a time step of 0.0015 ps for 150 ps (in total $10^5$ time steps). After the sample attains thermal equilibrium, the rigid asperity is moved along the $y$ direction with constant velocity $V$ ranging from 3.6 m/s to 360 m/s (these velocities approximately correspond to 0.01$a$/ps to 1.0$a$/ps, with *a* being the lattice constant) with the different interference depth *h* (i.e., the asperity overlap). A canonical ensemble NVT (constant number of atoms, volume and temperature) is imposed where a Nosé-Hoover thermostat is used to maintain the system at a constant average temperature of 0.01 K, so that thermal effects are effectively eliminated. This allows us to focus solely on the effect of the plowing depth and loading velocity on the asperity plasticity. In the analysis, dislocation tracking is performed using the dislocation extraction algorithm (DXA) [38], and Ovito [39] is used to visualize the defect structures.

During the asperity plowing process, the normal (along the *z* direction) and tangential (along the *y* direction) interaction forces between the asperities are recorded. The tangential force, opposite to the sliding direction, and the compression load along the *z* direction are defined as positive. These forces, $F_y$ and $F_z$, are averaged over the whole plowing process, as explained in Eq. (1), and they are rewritten as:

$$\bar{F}_y = \frac{1}{t_1-t_0}\int_{t_0}^{t_1} F_y dt, \quad \bar{F}_z = \frac{1}{t_1-t_0}\int_{t_0}^{t_1} F_z dt. \tag{3}$$

Here, $t_0$ and $t_1$ are the moments when the asperities come in contact and lose contact, respectively. For asperities of different sizes, normalized forces are also introduced as follows:

$$\bar{F}_{ny} = \frac{\bar{F}_y}{ER^2}, \quad \bar{F}_{nz} = \frac{\bar{F}_z}{ER^2}, \tag{4}$$

where *E* is the Young's modulus of copper and is assumed to be 120 GPa in this study.

## 3. Results and analysis of a single plowed asperity

In this section, we focus on the single asperity mechanical response under different conditions, such as the interference depth, asperity size, and plowing velocity. The mechanical response of the asperity is then analyzed based on the evolution of the dislocation microstructure and atomic deformation.

### 3.1 Effects of the interference on asperity plowing process

The effect of the interference depth on the mechanical response of the asperity was studied during the single asperity plowing process. The deformable asperity is orientated as *x*[100], *y*[010], *z*[001]. The interference depth *h* is normalized by the asperity radius *R*: $\alpha = h/R$, and the dimensionless overlap *α* is in the range from 0.05 to 0.8. In Figure 3 (a) and (b), the normal and tangential forces are plotted as a function of the plowing distance $S_y$ normalized by the asperity size. The plowing velocity is set to 36 m/s. For the asperity of radius $30a$, the normal force, which is shown in Figure 3(a), increases initially with an increasing plowing distance, and then decreases; eventually, it becomes zero when asperities lose contact. In general, a larger interference depth results in a higher peak normal force. For $\alpha = 0.5$, the dislocation microstructures and asperity deformations at different plowing distances are illustrated in Figure 3(c). It can be seen that during the plowing process, complex and dense dislocation microstructures are formed, and the asperity is plastically deformed.

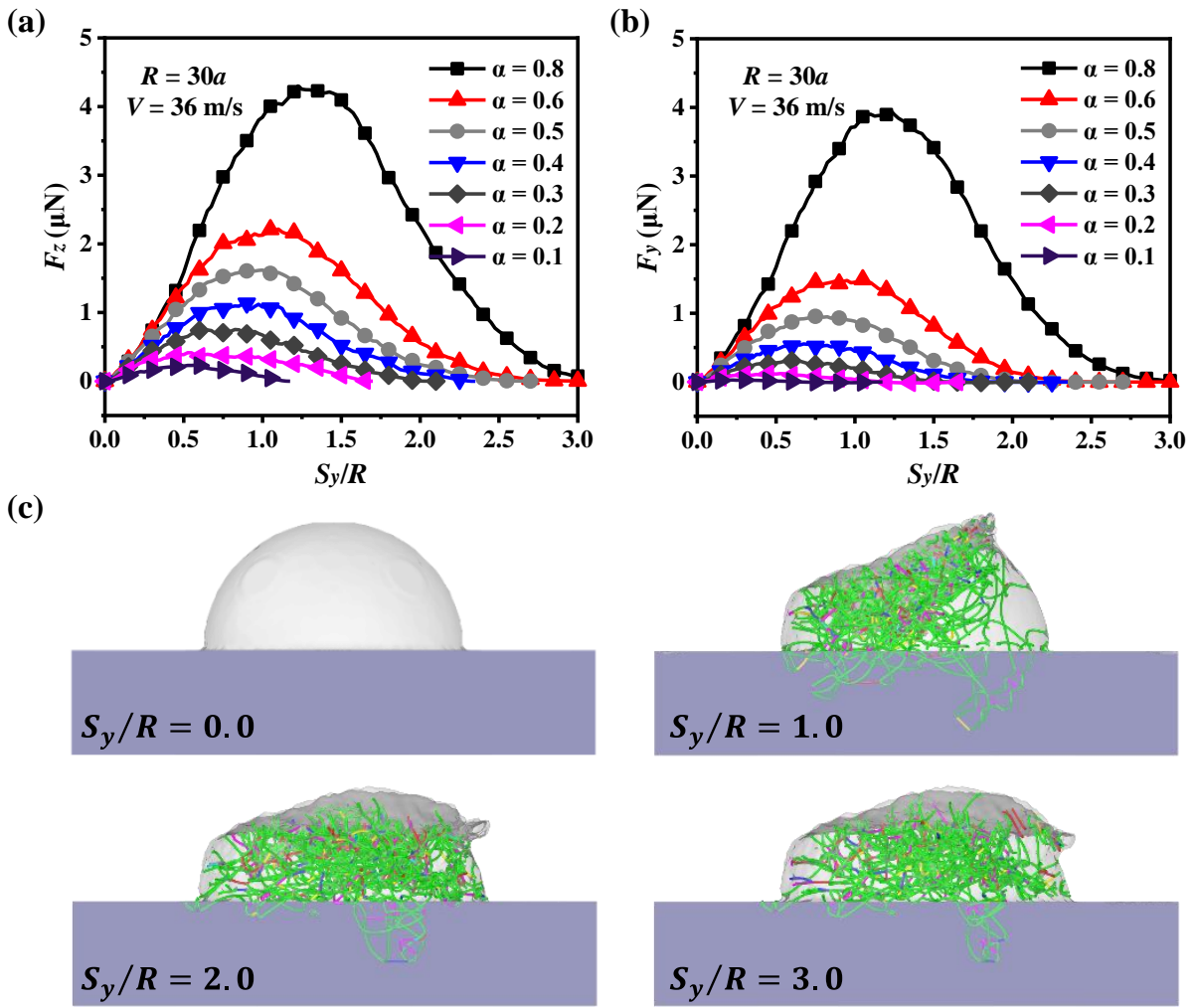

Figure 3: Asperity responses as a function of plowing distance. Marker symbols do not indicate data points and are only used to distinguish the lines easily. (a) Normal force $F_z$, (b) tangential force $F_y$. (c) Evolution of dislocation microstructures when $\alpha = 0.5$. The color scheme of the dislocation type follows Figure 6(b).

Tangential forces under different interference are shown in Figure 3(b), which exhibit similar features except for the one at a small interference depth. This can be seen in Figure 4(a) for $\alpha = 0.05$; the tangential force exhibits anti-symmetry across the entire plowing distance. The main reason is that the contact interface is essentially frictionless, and there is not enough plasticity generated in the deformable asperity. The deformation process is mostly elastic, which has an antisymmetric feature for the tangential force. The dislocation microstructures at peak forces are also shown in Figure 4(b). There are dislocations that are nucleated; however, these dislocations cannot glide into the asperities owing to the image force from the surface. They escape through the surface soon after they are generated; therefore, deformable asperity is essentially elastic.

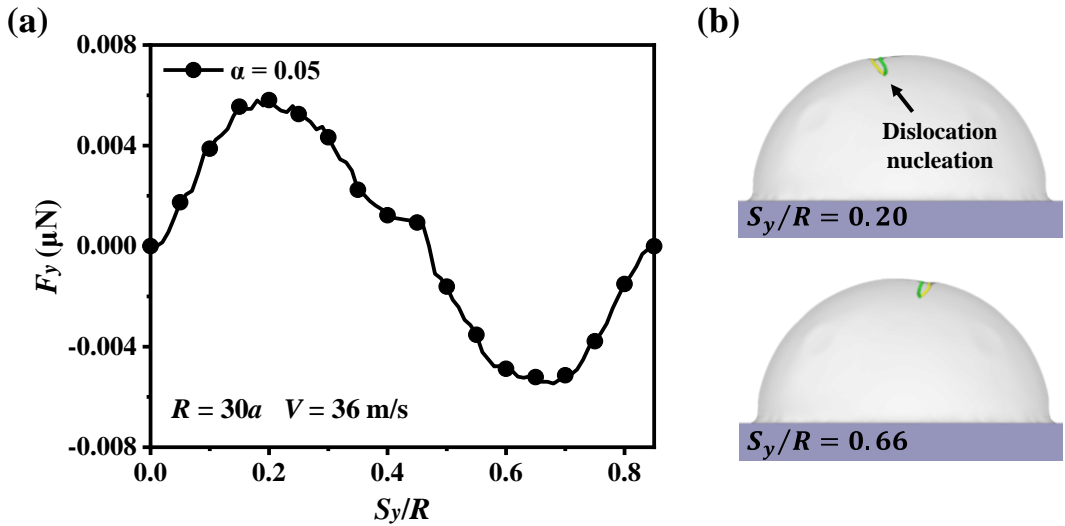

Figure 4: (a) Evolution of tangential force when $\alpha = 0.05$. (b) Dislocation microstructures at peak forces, and the corresponding plowing distance are also shown.

### 3.2 Effect of the asperity size

We now discuss the effect of the asperity size on its mechanical behavior under plowing for a plowing depth $\alpha = 0.5$. As shown in Figure 5(a) and (b), normal and tangential forces during the plowing process increase as the asperity size increases. This is expected because more atoms are in contact with the indenter for larger asperity. The evolution of the forces shows the same trend, namely, as the plowing distance increases, the two forces first increase and then decrease until the asperities are no longer in contact. The normalized forces $F_{nz}$ and $F_{ny}$, which are defined in Eq. (4), are also shown in Figure 5(c) and (d). It can be seen that the normal and tangential forces generally exhibit size-dependent behavior; a larger asperity results in a larger normalized force. Furthermore, this effect is more pronounced for the tangential force.

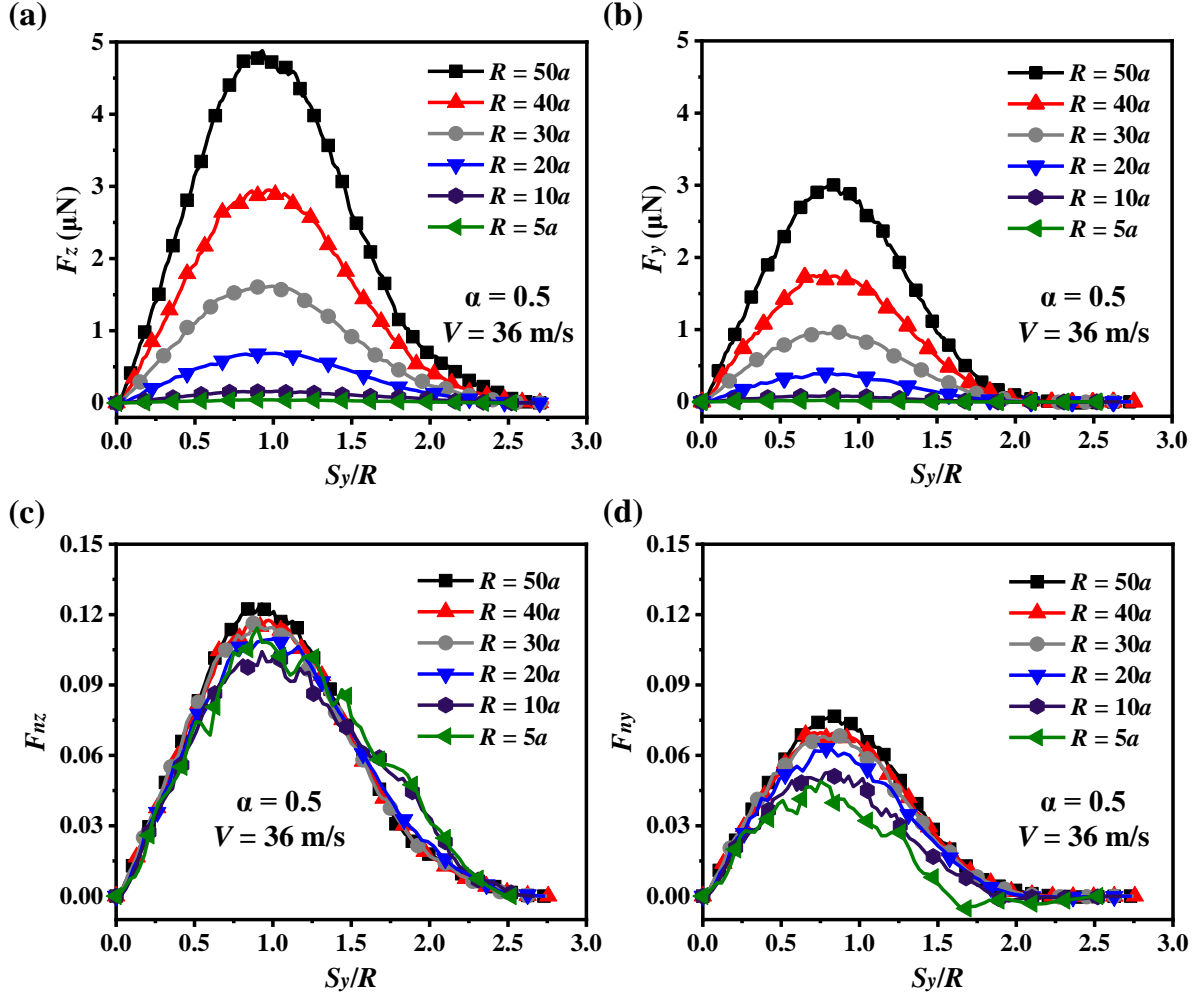

Figure 5: Effect of asperity size on asperity response when the dimensionless plowing depth $\alpha$ = 0.5. (a) Normal force, (b) tangential force; (c) Normalized normal force, and (d) normalized tangential force.

Typical asperity deformation and dislocation microstructures for various asperity sizes after the plowing process are shown in Figure 6(b). It can be seen that for a very small asperity ($R = 5a$), after deformation, the asperity is dislocation free. For larger asperities (e.g., $R = 30a$), complex dislocation microstructures (mainly Shockley partial dislocations) are left in the asperity. The dislocation microstructures in bigger asperities appear to fill the entire asperity. The total length of dislocations can be calculated using DXA [38]. With this, the dislocation density is estimated by $\rho = l/V_a$, where $l$ is the total length of dislocations, and $V_a = 2\pi R^3/3$ is the initial volume of the asperity, as shown in Figure 6(a). For asperities with radius $R \geq 20a$, the figure shows that the dislocation density is approximately constant.

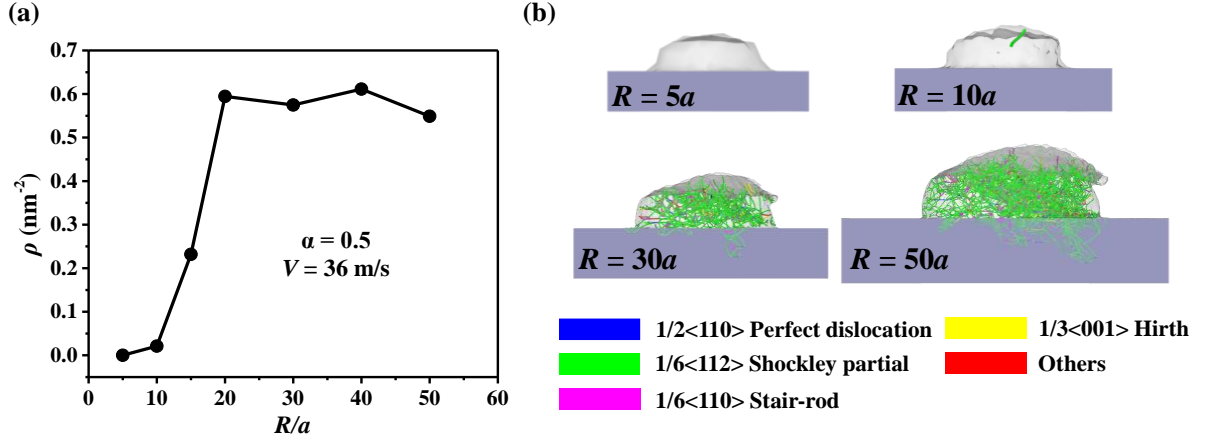

Figure 6: (a) The dislocation density and (b) the dislocation microstructures in asperities of different sizes when the asperities are out of contact.

Following the general idea schematically shown in Figure 1, we also calculated the time-averaged asperity strength during plowing: the normal and tangential forces during the plowing process, shown in Figure 5, are time-averaged following Eq. (3) and then normalized using Eq. (4). Figure 7 shows that the time-averaged asperity strength, including both normal and tangential directions, becomes almost size independent with the increase in the asperity size. The physical interpretation of this phenomenon is that for the same dimensionless plowing depth α, the average strength of a "statistical asperity ensemble" (i.e., the average strength of asperities with all possible contact configurations) becomes size independent. If one defines the friction coefficient as the ratio of the tangential strength to the normal strength, in our simulation system, the friction coefficient exhibits size-dependent behaviors when the asperity size is smaller than 7.2 nm ($20a$). Hereafter, we decide to only consider large asperities because for them, the effect of size is not important, and this enables the study of other features/aspects, such as surface roughness and loading velocity in a cleaner manner.

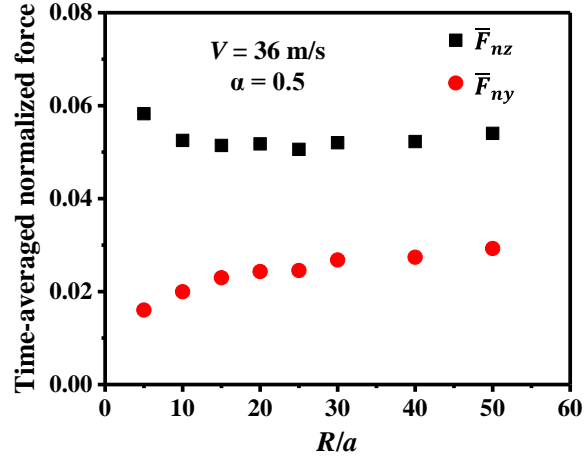

Figure 7: Variation in time-averaged asperity strength as a function of asperity size.

### 3.3 Effect of the plowing velocity

Subsequently, we focus on the effect of the plowing velocity on the mechanical responses of the asperity. The normal and tangential forces for different velocities are shown in Figure 8. Two clear features are observed. First, with the increase in the plowing velocity, the forces increase rapidly. For instance, the peak force in the normal direction increases from 0.69 μN to 90.7 μN when the plowing velocity changes from 7.2 m/s to 360 m/s. Second, for a low plowing velocity (e.g., $V = 7.2$ m/s), the peak force appears earlier rather than in the middle of the whole process.

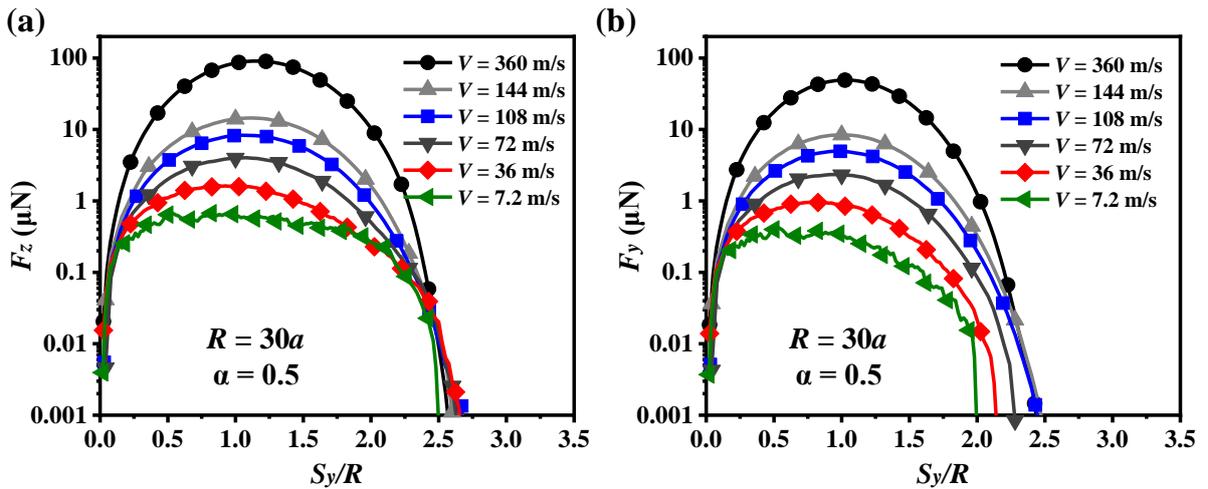

Figure 8: Effect of the plowing velocity on the asperity response for a dimensionless overlap coefficient $\alpha = 0.5$ and the asperity size $R = 30a$. (a) Normal force, (b) tangential force.

The effect of the plowing velocity on the dislocation microstructure evolution is shown in Figure 9 where the plowing distance $S_y$ is normalized by the total plowing displacement $S$ of the overall plowing process. Two clear features can be observed:

1) With increasing plowing velocity, the dislocation microstructure becomes denser, as confirmed by the evolution of the dislocation density in Figure 9(a). In single crystals, two main mechanisms are responsible for the stress relaxation in the system through plastic deformation: dislocation nucleation and dislocation propagation; the dominant mechanism depends on the external loading rate [40]. When the external loading velocity is small, the dislocations get nucleated and then propagate, having enough time to relax the stress of the system. As a consequence, the dislocation microstructure is more localized because the activated slip systems are sufficient to relax the stress. However, when the loading rate/velocity is high, nucleated dislocations either do not have enough time to propagate or the plasticity generated by dislocation propagation is not sufficient to relax the high internal stress. In the latter case, further nucleation on other slip systems is triggered, which eventually results in a denser dislocation microstructure [40].

2) The material pile up at the right side of the asperity grows with increasing plowing velocity. This is caused by massive dislocation nucleation in the asperity and subsequent dislocation annihilation at the surface. The shear stress distribution in the asperity caused by plowing has a high value front ahead of the rigid asperity (Ref. Figure 3(b) in [19] for the two dimensional (2D) plane strain case). Under high plowing velocity, because the stress cannot be effectively relaxed, the high-stress region would continuously nucleate dislocations, which in turn continuously annihilate at the free surface and leave a surface step. These surface steps accumulate to form a material pile up, which would eventually turn into wear debris. This phenomenon may also explain the increasing wear rate with increasing sliding velocity [41] for metal surfaces.

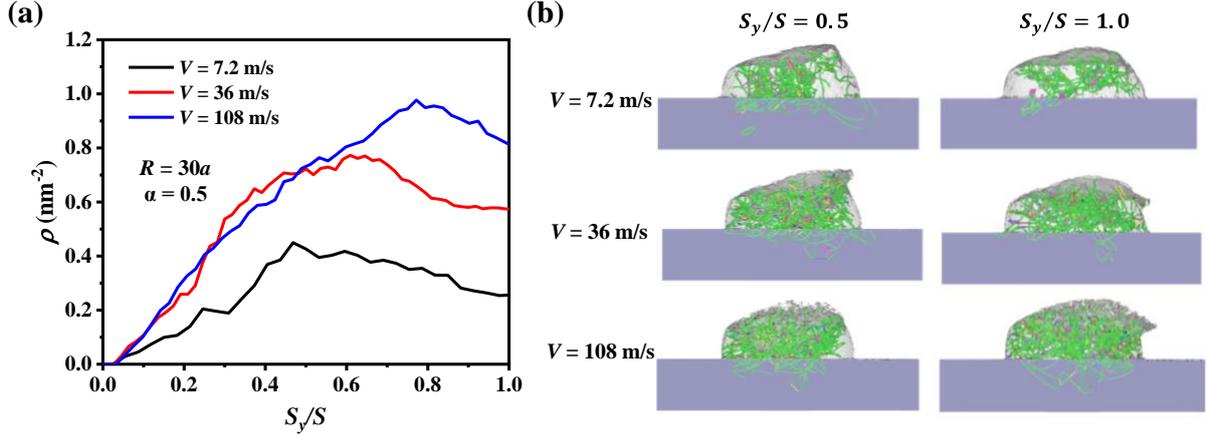

Figure 9: (a) Dislocation density as a function of plowing distance and (b) dislocation microstructures in the asperity at different plowing velocities when $\alpha = 0.5$. Color scheme of dislocation type follows that of Figure 6(b).

The time-averaged normalized forces for different plowing velocities are shown in Figure 10. It can be seen that the normalized forces can be clearly divided into two regimes distinguished by the different plowing velocities. The distinction between the two regimes is made through a quantity $\eta = (\bar{F}_{ny} - \bar{F}_{ny0})/\bar{F}_{ny0}$, where $\bar{F}_{ny}$ and $\bar{F}_{ny0}$ are the time-averaged-normalized tangential forces corresponding to a given velocity $V$ and an extremely low plowing velocity $V_0$, respectively. In this study, $V_0$ is taken as 3.6 m/s. The mechanical response is considered to be low-velocity dependent when $\eta \leq 20\%$. As shown in Figure 10(a), when the asperity interference is $\alpha = 0.5$, for the low plowing velocity (e.g., $V < \sim 7.2$ m/s), asperity responses exhibit low-velocity dependence. When the plowing velocity exceeds a certain value, which is appropriately 12 m/s, the normalized forces start to show a strong velocity dependence. The critical/transition plowing velocity also depends on the interference depth, as shown in Figure 10(b), where $\alpha = 0.2$, and the transition velocity is approximately 50 m/s.

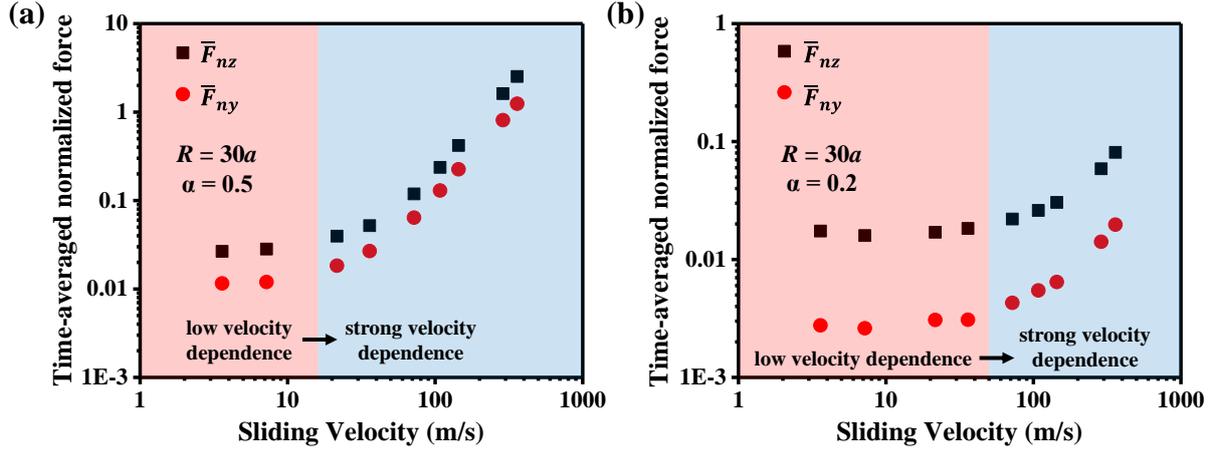

Figure 10: Time-averaged normalized forces as a function of plowing velocity, asperity radius $R = 30a$. Two different interference depths are studied (a) $\alpha = 0.5$ and (b) $\alpha = 0.2$.

## 4. Statistical model for slides of macroscopic rough surfaces

In the previous section, we focused on the mechanical response of a single asperity under different loading conditions. In this section, we aim to utilize the single asperity response to predict the rough surface response. Macroscopic rough surfaces consist of various asperities of different heights, and the surface property, such as COF, is the integrand of all asperity pairs. The GW-type model [2] is introduced here to study the COF. As shown in Figure 11, we consider rigid wavy rough surface slides over a deformable rough surface whose asperity height follows a Gaussian distribution:

$$\psi(z) = \frac{1}{\sigma\sqrt{2\pi}} \exp\left(\frac{-z^2}{2\sigma^2}\right). \tag{5}$$

Here, $\sigma$ is the standard deviation of the asperity heights. The distance between the asperity peak on the rigid wavy surface and the mean asperity height of the deformation surface is $d$, which is determined by the normal force that puts two surfaces into contact before sliding. A larger normal force results in a smaller $d$.

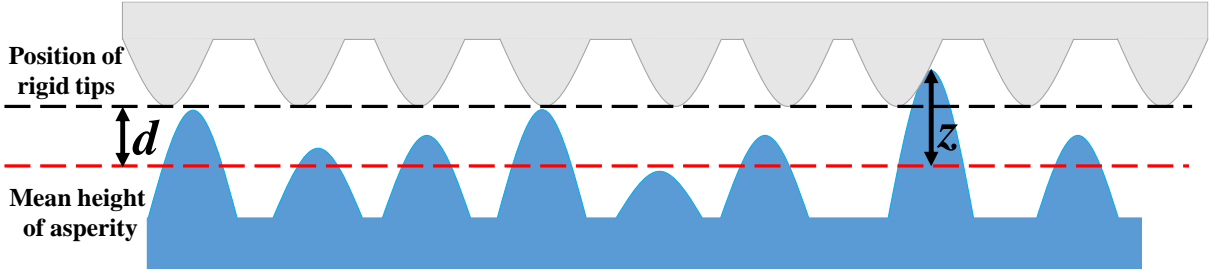

Figure 11: Schematic of macroscopic contact between rigid surface (gray) and deformation surface (light blue).

To obtain the macroscopic contact forces (i.e., the normal and tangential forces), we assume that all asperities deform independently. The plowing/interference depth for an asperity of height $z$ is $z$-$d$ when $z > d$; therefore, for the macroscopic contact with $N_0$ asperities on each surface and the total normalized (normalized by $ER^2$ as shown in Eq. (4)) tangential and normal forces between the contact surfaces during sliding are:

$$F_{ts} = N_0 \int_d^{+\infty} \bar{F}_{ny} \psi(z) dz. \qquad (6)$$

$$F_{ns} = N_0 \int_d^{+\infty} \bar{F}_{nz} \psi(z) dz. \qquad (7)$$

By substituting the single asperity response in the above equations, the COF can then be calculated by $F_{ts}/F_{ns}$ or $F_{ts}^*/F_{ns}^*$, where the macroscopic normalized tangential and normal forces are $F_{ts}^* = F_{ts}/N_0$ and $F_{ns}^* = F_{ns}/N_0$, respectively. Here, we note that our study is multiscale in the sense that the mechanical response on the asperity level is integrated to obtain the mechanical response of the surface level. It can be seen that the statistical model requires a single asperity response to be a function of the asperity height $z$ (alternatively $\alpha$, since $\alpha = (z-d)/R$); therefore, we have carried out a large number of simulations to fit the functional forms for single asperity responses. Details can be found in Appendix B. Because the single asperity responses also strongly depend on the crystal orientation, we have also studied the orientation effect. Details can be found in Appendix A. Overall, the functional forms are fitted for three different loading velocities (when the crystal orientations were the same), for different crystal orientations (under the same loading velocity). This enables us to study the effect of loading (sliding) velocity and crystal orientation on the friction coefficient using a statistical model.

As shown in Figure 12(a), both the normal and tangential forces depend on the surface separation distance $d$: a smaller $d$ results in larger forces. It can also be seen that both forces depend on the surface roughness (which is defined as $\sigma/R$): a rougher surface will lead to larger forces. Furthermore, in the inset, we show the sensitivity of the forces with respect to $d$. It can be clearly seen that the tangential force is more sensitive to $d$ than the normal force. In Figure 12(b), the COFs of different surface roughness are plotted against the change in $d$ between contact surfaces. Under a specific sliding velocity of 36 m/s, the increase in surface roughness results in a larger COF. It is interesting to note that decreasing $d$ will increase the COF, which is consistent with the sensitivity shown in Figure 12(a) inset. This appears to contradict Amonton's third law and macroscopic experimental measurements, where the COF is believed to be a constant that only depends on the material property and surface roughness. However, it has been observed experimentally [42] that the nanoscale friction coefficient resulting from plasticity depends on the normal force; a larger normal force (i.e., smaller $d$) results in a higher COF.

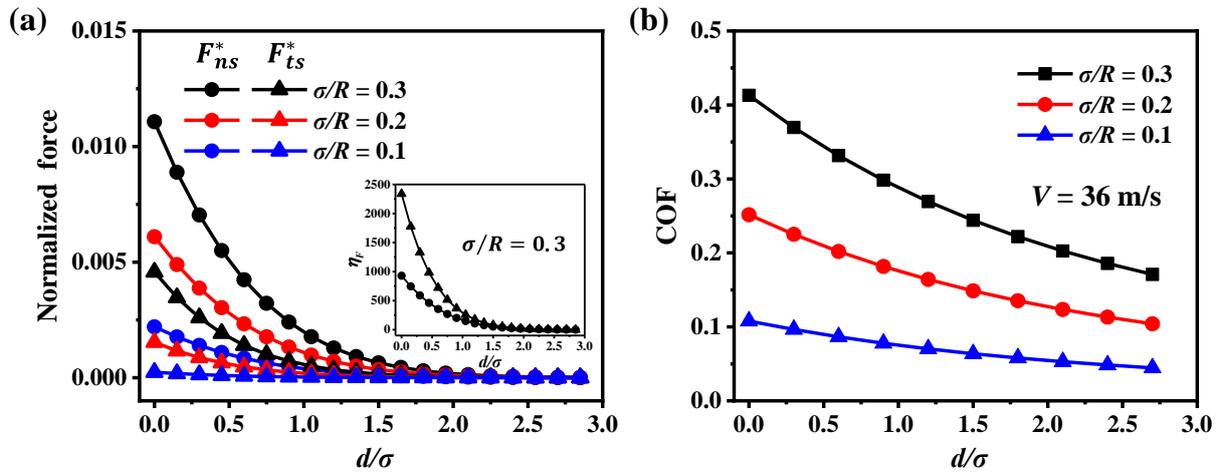

Figure 12: Effect of the surface roughness on (a) normalized force and (b) COF as a function of $d$. The inset in (a) shows the forces normalized by their corresponding values at $d/\sigma = 2.85$, which is difficult to observe in the main figure of (a). The ratios show the unbiased sensitivity.

It can be seen in Figure 13(a) that the COF evaluated by the statistical model exhibits velocity dependence, which is consistent with the experimental observation: the COF generally increases with an increasing sliding velocity [43]. At the same time, it is seen that the velocity dependence in our model varies with *d*. When *d* is large (i.e., the normal contact force is small), most of the asperity plowing pairs will have small interference. As shown in Figure 10(b), for small interference, the critical transition velocity is high; therefore, 36 m/s and 7.2 m/s exhibit weak velocity dependence. Similarly, when *d* is small, there will be more asperity pairs with high values of interference, where the critical transition velocity is low, as shown in Figure 10(a); therefore, the COF exhibits strong velocity dependence. It should be noted that the velocity-dependent COF here mainly originates from rate-dependent dislocation plasticity, rather than other mechanisms in other frictional systems. However, the friction coefficient in our system can also be described well using the commonly used rate and state friction framework, whose general form is $\mu = \mu_0 + (a - b)\ln(V/V_0)$ [44, 45]. In this framework, the velocity dependence is described by parameter (*a-b*), which needs to be measured experimentally for the specific testing system. In our model, the parameter (*a-b*) is determined by *d*.

Figure 13(b) shows the dependence of the COF on the crystal orientation. The results obtained confirm the estimation made based on the single asperity response in Appendix A; the friction coefficient is smaller along the atom close-packed direction than of those in other directions.

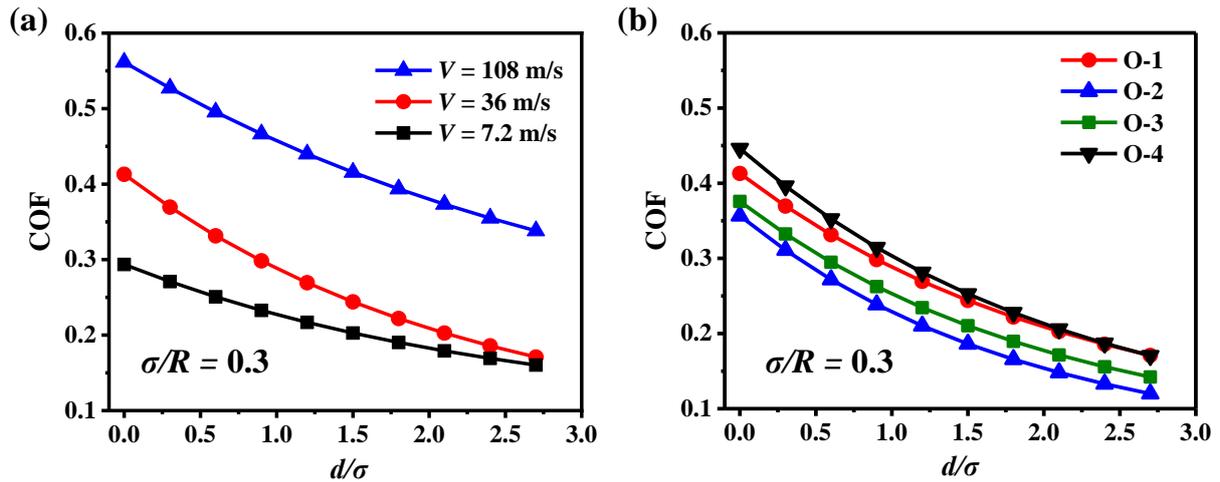

Figure 13: Effects of (a) sliding velocity and (b) crystal orientation on COF as a function of *d*.

## 5. Summary and conclusions

In this study, we used a GW-type friction model to create a multiscale model for obtaining the macro-frictional behavior of rough surfaces from the plowing response of a single asperity at the nanoscale.

MD simulations are performed to investigate the non-adhesive plowing of a single asperity. The effects of interference depth, asperity size, relative plowing velocity, and crystal orientation are discussed as follows: 1) the friction forces (frictional strength) are size dependent, and the friction coefficient increases with increasing asperity size. However, it becomes insensitive to the asperity size when the asperity is larger than a critical size (radius $R > \sim7.2$ nm); (2) we find a critical plowing velocity below which the asperity responses exhibit weak velocity dependence; (3) when the plowing direction is parallel to the crystal close-packed plane, it is easy to plow the asperity.

Using the statistical model, we studied the frictional behavior of rough surfaces. The salient conclusions are as follows:

1. Smaller surface separation $d$ yields a higher COF, which contradicts Amonton's law, but is consistent with nanoscale experimental observations.

2. The COF increases with the increasing surface roughness. When the surface is rougher, the COF is more sensitive to $d$.

3. A higher sliding velocity results in a higher COF. At the same time, the dependence on the sliding velocity strongly relies on the surface separation distance $d$, which can be incorporated into the rate and state friction framework.

4. The crystal orientation has a clear influence on the COF; when the atomic close-packed surface is parallel to the sliding direction, the COF is smaller.

**Acknowledgments**

This work was supported by the National Natural Science Foundation of China (11802310, 11772334), the Youth Innovation Promotion Association CAS (2018022), and by the Strategic Priority Research Program of the Chinese Academy of Sciences (No. XDB22040501). HS and SS acknowledge financial support from the European Research Council through the ERC Grant Agreement No. 759419 MuDiLingo ("A Multiscale Dislocation Language for Data-Driven Materials Science").

**Supplementary Material**

The data shown in the figures along with the python plot scripts are available at https://gitlab.com/computational-materials-science/public/publication-data-and-code/2020_Hu-et-al__MultiscaleStudyOfTheDynamicFrictionCoefficient

**Appendix A: Effect of crystalline orientation**

Owing to the anisotropic mechanical behavior of crystalline materials at the atomic scale, the frictional properties of the material also strongly depend on the crystal orientation [46-48]. In this appendix, asperities with four different crystal orientations, as listed in Table A1, were studied to reveal the underlying orientation dependence. The asperity size is chosen as $30a$ (i.e., $R = \sim10.8$ nm) and the plowing velocity is 36 m/s along the $y$ direction. The dimensionless interference depth $\alpha$ ranges from 0.1 to 0.8 (only two typical overlaps are shown in Figure A1). With increasing overlap, the mechanical response of the asperities exhibits significant orientation dependence. For asperity contact with a large overlap (i.e., $\alpha = 0.5$), the tangential forces in the asperities with $z[111]$ (orientations O-2 and O-3) are relatively smaller than the other orientations (O-1 and O-4), as shown in Figure A1(b). The asperities with orientation O-2 have a higher normal force than those of the orientation O-3 in the case of both large and small overlaps, as shown in Figure A1(a).

Table A1 Details of asperity orientations.

| Orientation Number | Crystal Orientation |
|---|---|
| O-1 | $x[100]$, $y[010]$, $z[001]$ |
| O-2 | $x[1\bar{1}0]$, $y[11\bar{2}]$, $z[111]$ |
| O-3 | $x[\bar{1}\bar{1}2]$, $y[1\bar{1}0]$, $z[111]$ |
| O-4 | $x[1\bar{1}0]$, $y[11\bar{1}]$, $z[112]$ |

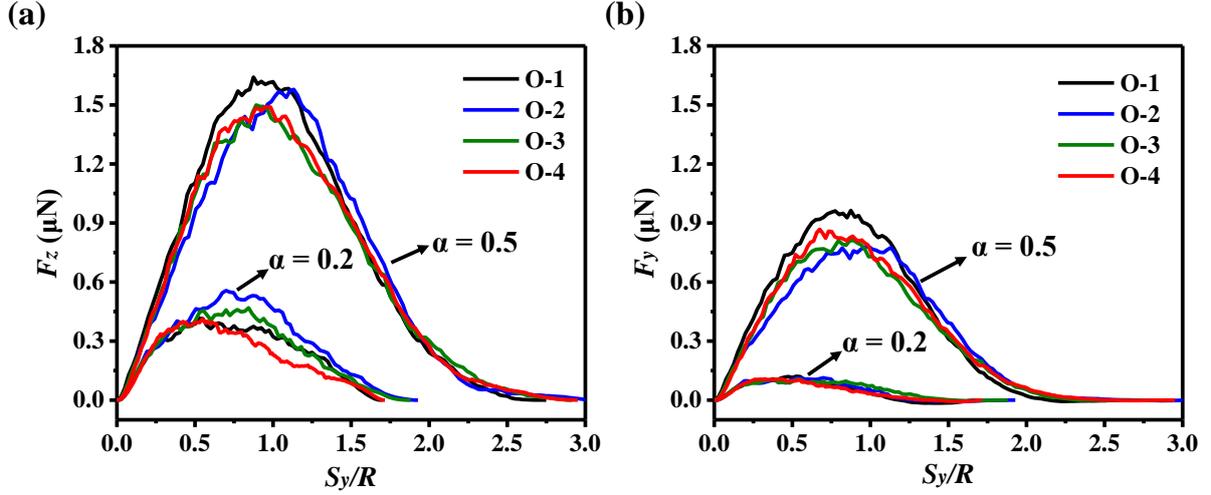

Figure A1: Responses of asperities with different crystal orientations: (a) Normal force and (b) tangential force.

As shown in Figure A2, the distributions of strain $\varepsilon_{yz}$ as well as the associated dislocation microstructures at a specific interference depth $\alpha = 0.5$ when $S_y/S$ is approximately 0.1 ($F_y:F_z = 1:2$) are analyzed to reveal the orientation-dependent deformation mechanism. For asperities with O-2 and O-3, the plowing directions are parallel to the atomic close-packed surfaces (i.e., $z$[111]), where dislocations are easier to get nucleated; therefore, these two orientations are expected to have smaller friction coefficients (because it is easier to deform in the tangential direction) when compared to those of other orientations. However, the dislocation structures are also quite different in these two orientations, as shown in Figure A2(b). Because most of the activated dislocations in Figure A2(b) are Shockley partial dislocations, we calculated the Schmid factors on the slip systems 1/6<112>{111} for the four crystal orientations. The calculation of the Schmid factor considers the ratios of the tangential force to the normal force (i.e., $F_y:F_z$) to be between 1:1 and 1:3 based on the simulation results. The active slip planes are summarized in Table A2. It can be seen that the active slip planes in orientation O-3 are ($\bar{1}11$) whose deformation contributes to the normal load direction, that is, the normal load would be released, so it is expected that the O-3 orientation has a larger friction coefficient than the O-2 orientation. For the other two orientations whose plowing directions are not parallel to the close-packed plane, the active slip planes for O-4 are mostly (111) during the loading process, in contrast to O-1 whose active slip planes have a normal component; therefore, O-4 is expected to have a larger friction coefficient than O-1. Even though the above analysis does not consider the effect of the dislocation microstructure evolution during the deformation, the friction coefficients in the four orientations can be roughly estimated as O-4 > O-1 > O-3 > O-2.

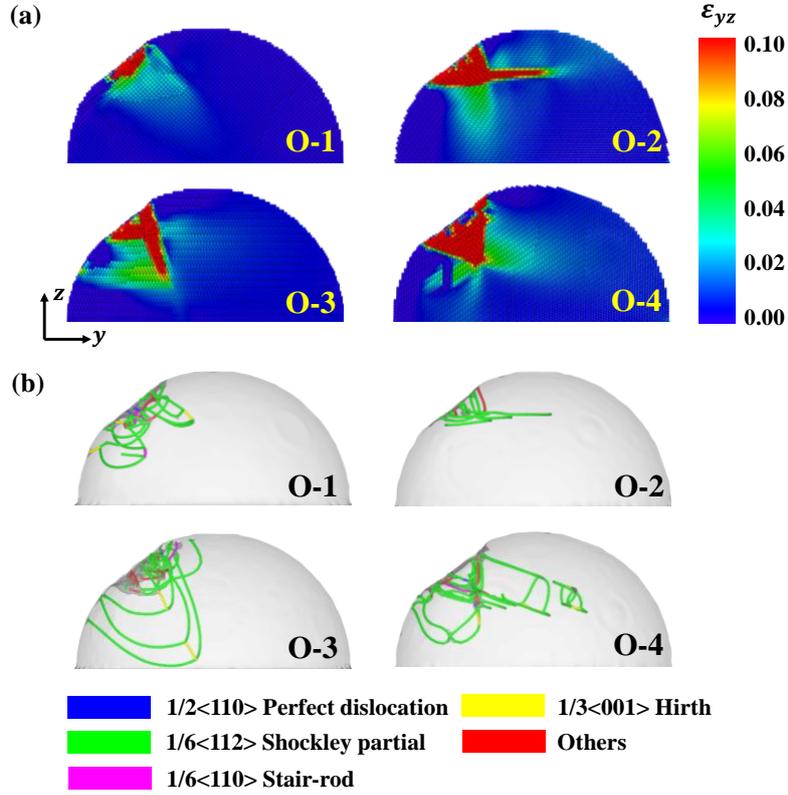

Figure A2: (a) Distribution of strain $\varepsilon_{yz}$ and (b) dislocation structures in the asperities of different orientations when $\alpha = 0.5$ and $S_y/S = \sim 0.1$.

Table A2 Activated slip planes under different external loads.

| Orientation Number | $F_y:F_z = 1:1$ | $F_y:F_z = 1:2$ | $F_y:F_z = 1:3$ |
|---|---|---|---|
| O-1 | $(1\bar{1}1), (11\bar{1})$ | $(1\bar{1}1), (11\bar{1})$ | $(1\bar{1}1), (11\bar{1})$ |
| O-2 | $(111)$ | $(111)$ | $(\bar{1}11), (1\bar{1}1)$ |
| O-3 | $(\bar{1}11)$ | $(\bar{1}11)$ | $(\bar{1}11)$ |
| O-4 | $(11\bar{1})$ | $(111)$ | $(111)$ |

**Appendix B: Fitting functions of single asperity mechanical responses**

Macroscopic contacts contain numerous micro and nanoasperity pairs at small length scales. When the two surfaces in contact slide relative to each other, asperity pairs of different sizes with different interference depths will move under the same sliding velocities. Thus, in order to obtain the surface response, it is necessary to quantitatively pre-acquire the mechanical responses of asperity pairs. In this appendix, we aim to provide fitting formulas for the asperity mechanical response-based analysis in the above subsections.

As shown in Figure B1(a) and (b), for the asperity of crystal orientation O-1, the time-averaged normalized average forces obtained by MD simulations are fitted in the power-law form as a function of the interference depth when the plowing velocity is 36 m/s. Similarly, the forces between asperities are also fitted for plowing velocities of 7.2 m/s and 108 m/s. The power-law fitting parameters at different plowing velocities are summarized in Table B1. Similarly, the time-averaged normalized average forces in the asperities of the other three orientations at a plowing velocity of 36 m/s are also fitted in the same way, and the fitting parameters are summarized in Table B2.

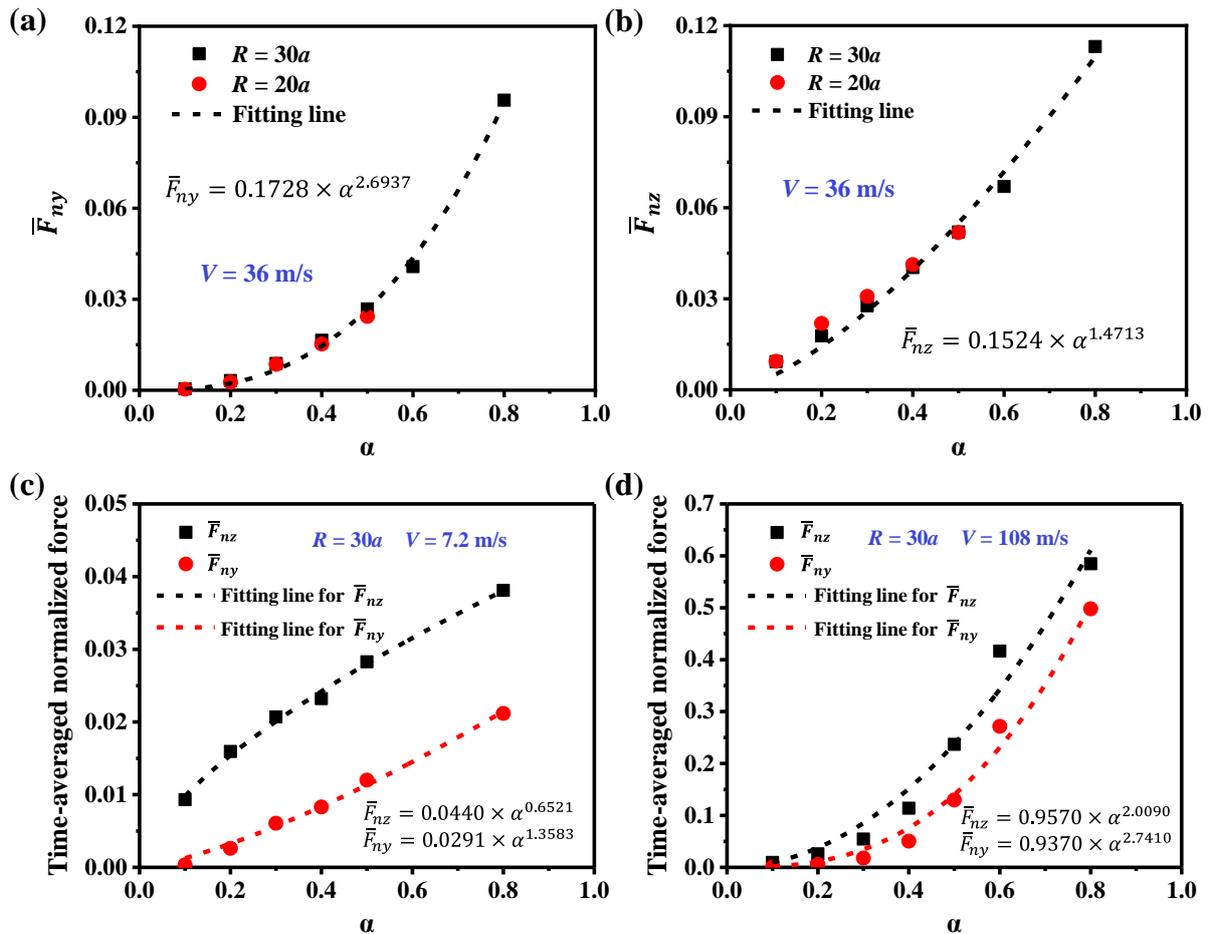

Figure B1: MD simulation results and data fitting for (a) time-averaged normalized tangential

force at $V$ = 36 m/s and (b) time-averaged normalized normal force at $V$ = 36 m/s; both the time-averaged normalized forces at (c) $V$ = 7.2 m/s and (d) $V$ = 108 m/s.

Table B1 Fitting parameters of power-law function ($\bar{F} = m \times \alpha^n$) for three different plowing velocities when the crystal orientation is O-1.

| Plowing velocity | | 7.2 m/s | 36 m/s | 108 m/s |
|---|---|---|---|---|
| $\bar{F}_{ny}$ | $m$ | 0.0291 | 0.1728 | 0.9370 |
| | $n$ | 1.3583 | 2.6937 | 2.7410 |
| $\bar{F}_{nz}$ | $m$ | 0.0440 | 0.1524 | 0.9570 |
| | $n$ | 0.6521 | 1.4713 | 2.0090 |

Table B2 Fitting parameters of power-law function ($\bar{F} = m \times \alpha^n$) for asperities of different crystal orientations at a plowing velocity of 36 m/s.

| Orientation Number | | O-2 | O-3 | O-4 |
|---|---|---|---|---|
| $\bar{F}_{ny}$ | $m$ | 0.1379 | 0.1331 | 0.2176 |
| | $n$ | 2.7862 | 2.6648 | 3.3246 |
| $\bar{F}_{nz}$ | $m$ | 0.1099 | 0.1148 | 0.1751 |
| | $n$ | 1.2889 | 1.3347 | 1.8985 |